\documentclass[aps,pra,floats,floatfix,onecolumn]{revtex4}

\usepackage{ifthen}
\usepackage{ifpdf}
\usepackage{color}

\ifpdf
\usepackage{graphicx}
\usepackage{epstopdf}
\else
\usepackage{graphicx}
\usepackage{epsfig}
\fi
\graphicspath{{../}}

\usepackage{latexsym}
\usepackage{amsmath}
\usepackage{amssymb}
\usepackage{bm}
\usepackage{wasysym}

\newcommand{\varphiJ}{\bm{\varphi}}

\newcommand{\nJ}{\bm{n}}

\newcommand{\const}{\mbox{const}}
\newcommand{\trc}{\mbox{trace}}

\newcommand{\bra}{\left\langle}
\newcommand{\ket}{\right\rangle}

\newcommand{\tbox}[1]{\mbox{\tiny #1}}
 
\newcommand{\be}[1]{\begin{eqnarray}\ifthenelse{#1=-1}{\nonumber}{\ifthenelse{#1=0}{}{\label{e#1}}}}
\newcommand{\beq}{\begin{eqnarray}}
\newcommand{\eeq}{\end{eqnarray}} 

\newcommand{\hide}[1]{\textcolor{red}{[hidden text]}}

\newcommand{\Eq}[1]{\textcolor{blue}{Eq.\!\!~(\ref{#1})}} 
\newcommand{\Fig}[1]{\textcolor{blue}{Fig.}\!\!~\ref{#1}}

\begin{document}



\title{Temporal quantum fluctuations in the fringe-visibility of atom interferometers  with interacting Bose-Einstein condensate}

\author{Doron Cohen$^{1}$ and Amichay Vardi$^{2}$}
\affiliation{
Departments of Physics$^1$ and Chemistry$^2$, Ben-Gurion University of the Negev, P.O.B. 653, Beer-Sheva 84105, Israel\\}

\begin{abstract}
We formulate a semiclassical approach to study the dynamics of coherence loss and revival in a Bose-Josephson dimer. The phase-space structure of the bi-modal system in the Rabi, Josephson, and Fock interaction regimes, is reviewed  and the prescription for its WKB quantization is specified. The local density of states (LDOS) is then deduced for any given preparation from its semiclassical projection onto the WKB eigenstates. The LDOS and the non-linear variation of its level-spacing are employed to construct the time evolution of the initial preparation and study the temporal fluctuations of interferometric fringe visibility. The qualitative behavior and characteristic timescales of these fluctuations are set by the pertinent participation number, quantifying the spectral content of the preparation. We employ this methodology to study the Josephson-regime coherence dynamics of several initial state preparations, including a Twin-Fock state and three different coherent states that we denote as 'Zero', 'Pi', and 'Edge' (the latter two are both on-separatrix preparations, while the Zero is the standard ground sate preparation). We find a remarkable agreement between the semiclassical predictions and numerical simulations of the full quantum dynamics. Consequently, a characteristic distinct behavior is implied for each of the different preparations. 

\end{abstract}

\maketitle

\section{Introduction} 

Atom interferometry \cite{Ramsey50,Borde89,Berman97,Cronin09} with Bose-condensed atoms offers the possibility of constructing compact and highly precise  measurement tools. Recent experiments \cite{Andrews97,Shin04,Wang05,Albiez05,Jo07,JoChoi07,JoChoi07b,Schumm05,Hofferberth07,Est08,Boehi09,Gross10,Riedel10,Chen11,Luke11,Bucker11} demonstrate that bi-modal Bose-Einstein condensates (BECs) have the necessary phase-coherence and controllability of coupling and interaction parameters, to operate atom interferometers at the best sensitivity allowed by quantum mechanics.

A typical atom interferometer follows the Mach-Zehnder scheme consisting of a preparation stage, in which the bimodal input state is mixed by a (usually 50:50) beam-splitter, a waiting time ($t$) during which the system evolves and the two modes acquire a relative phase difference~$\varphi$, and a measurement stage where the two condensates are either released and allowed to interfere or are mixed again by a second beamsplitter. In the former case,  the accumulated relative phase $\varphi$ correlates with the location of interference fringes, whereas in the latter it is reflected in the final atom number difference. 

The single-particle coherence of the split BEC is characterized by the many-realizations fringe-visibility function $g_{12}^{(1)}(t)$. The useful timescale of an interferometric measurement is set by decoherence. For trapped BEC interferometers, an important source for the loss of single-particle coherence, is the phase-diffusion induced by nonlinear inter-particle interactions \cite{Javanainen97,Vardi01,Greiner02,Widera08,Grond10,Tikhonenkov10,Grond11,Jo07}. For uncoupled condensates starting from a coherent preparation, this process amounts to a simple Gaussian decay of $g_{12}^{(1)}(t)$ 
on a  time scale $t \sim (U\sqrt{N})^{-1}$, where $U$ is the interaction strength and~$N$ is the number of particles. However, most current atom-interferometer setups with trapped BECs operate in the Josephson interaction regime, where coupling while small, still affects the dynamics and generates richer coherence evolution which includes oscillations, fluctuations, and recurrences of $g_{12}^{(1)}(t)$ due to quantum collapse and revival \cite{Wright96,Will10,Boukobza09a}.  The evolution of fringe visibility in the Josephson regime, thus offers a unique opportunity for the controlled study of strong correlation dynamics, beyond the usual focus on ground state properties.

Our objective here is to characterize the quantum dynamics of single-particle coherence and explore the dependence of its fluctuations on the initial preparation \cite{Boukobza09a,Boukobza09b,SmithMannschott09,Chuchem10}.   
For this purpose we employ a semiclassical picture of the quantum two-mode Bose-Hubbard model normally used to describe BEC atom interferometers.  In Section II We introduce the model Hamiltonian for $N$ boson in a dimer system. In section III we provide the prescription for the WKB quantization of the associated spherical phase space. Several experimentally viable preparation are introduced in Section IV, and their local density of states (LDOS) is characterized using a participation number~$M$, with marked differences in the dependence of~$M$ on the particle number~$N$, and on the interaction~$U$. Consequently we are able to analyze the observed temporal fluctuations in Section~V and determine their long time average and their characteristic variance. 
We show that the RMS of the fluctuations scales differently with~$N$, depending on the nature of the prepared state.

\section{Model Hamiltonian and phase-space structure} 

Matter-wave interferometers can be realized using double-well spatial confinement \cite{Andrews97,Shin04,Wang05,Albiez05,Jo07,JoChoi07,JoChoi07b,Schumm05,Hofferberth07,Est08} or internal spin states \cite{Boehi09,Gross10,Riedel10,Chen11,Luke11}. In both cases, the bimodal system of interacting atoms  is described to good accuracy \cite{Li09} by the tight-binding Bose-Hubbard Hamiltonian (BHH) \cite{Kitagawa93,Holland93,Pezze09,Grond10,Tikhonenkov10,Grond11}. Here we refer to $N$ particles in a two-site (bi-modal) system, also known as a 'dimer':
\beq
\label{BHH}
\mathcal{H} \ = \ 
\sum_{i=1,2}\left[ \mathcal{E}_i \hat n_i
+ \frac{U}{2} \hat n_i(\hat n_i-1)\right] 
- \frac{K}{2}(\hat a_2^{\dag}\hat a_1+ \hat a_1^{\dag}\hat a_2 )~,
\eeq
where $K$ is the hopping amplitude, $U$ is the interaction, and $\mathcal{E}_i$ are the on-site energies. One may define an SU(2) algebra, where 
\be{11}
{{J}}_z \ \ &\equiv& \ \ \frac{1}{2}(\nJ_1-\nJ_2) \ \ \equiv \ \ \nJ, \\
{{J}}_+ \ \ &\equiv& \ \ \hat{a}_1^{\dag}\hat{a}_2~,
\eeq
and $J_{-} =[J_{+}]^{\dag}$. Hence ${{J}}_x=({{J}}_+ + {{J}}_-)/2$ and ${{J}}_y=({{J}}_+ - {{J}}_-)/2i$. 
Rewriting the Hamiltonian (\ref{BHH}) in terms of these SU(2) generators, we see it is formally the same as that of a spin $j=N/2$ system, 
\beq
\label{spin}
\mathcal{H} \ = \ - \mathcal{E} \hat J_z+ U \hat J_z^2 - K \hat J_x  + \mbox{const}~,
\eeq
where ${\mathcal{E}=\mathcal{E}_2-\mathcal{E}_1}$ is the bias in the on-site potentials. 
It is thus clear that the characteristic dimensionless parameters which determine both stationary and dynamic properties, are 
\beq
u &\equiv&  NU/K \\
\varepsilon &\equiv& \mathcal{E}/{K}
\eeq 

The two-site BHH can be regarded as the quantized version of the top Hamiltonian, whose {\em spherical} phase space is described by the conjugate non-canonical coordinates ${(\bm{\theta},\bm{\varphi})}$ that are defined through
\beq
\hat J_z  &=&  [(j{+}1)j]^{1/2} \ \cos(\bm{\theta}),
\\
\hat J_x  &=&   [(j{+}1)j]^{1/2} \ \sin(\bm{\theta})\cos(\bm{\varphi}),  
\eeq
The Hamiltonian (\ref{spin}) is thus transformed into the top Hamiltonian,
\beq
\mathcal{H}(\bm{\theta},\bm{\varphi}) \ = \ 
\frac{NK}{2}\left[\frac{1}{2} u (\cos\bm{\theta})^2 - \varepsilon \cos\bm{\theta}  - \sin\bm{\theta}\cos\bm{\varphi} \right]~.
\label{top}
\eeq
which is closely related to the Josephson Hamiltonian.
The {\em cylindrical} phase space of the latter 
is described by the canonical coordinate $\bm{n}$ of (\ref{e11}),   
and its conjugate angle $\bm{\varphi}$. 
In the absence of bias, in the vicinity of the Equator, 
it is just the pendulum Hamiltonian:
\beq
\mathcal{H}(\bm{n},\bm{\varphi}) \ \approx \ 
U \bm{n}^2 - \frac{1}{2}KN \cos\bm{\varphi} 
\eeq

\begin{figure}[h!]

\centering
\includegraphics[width=7cm]{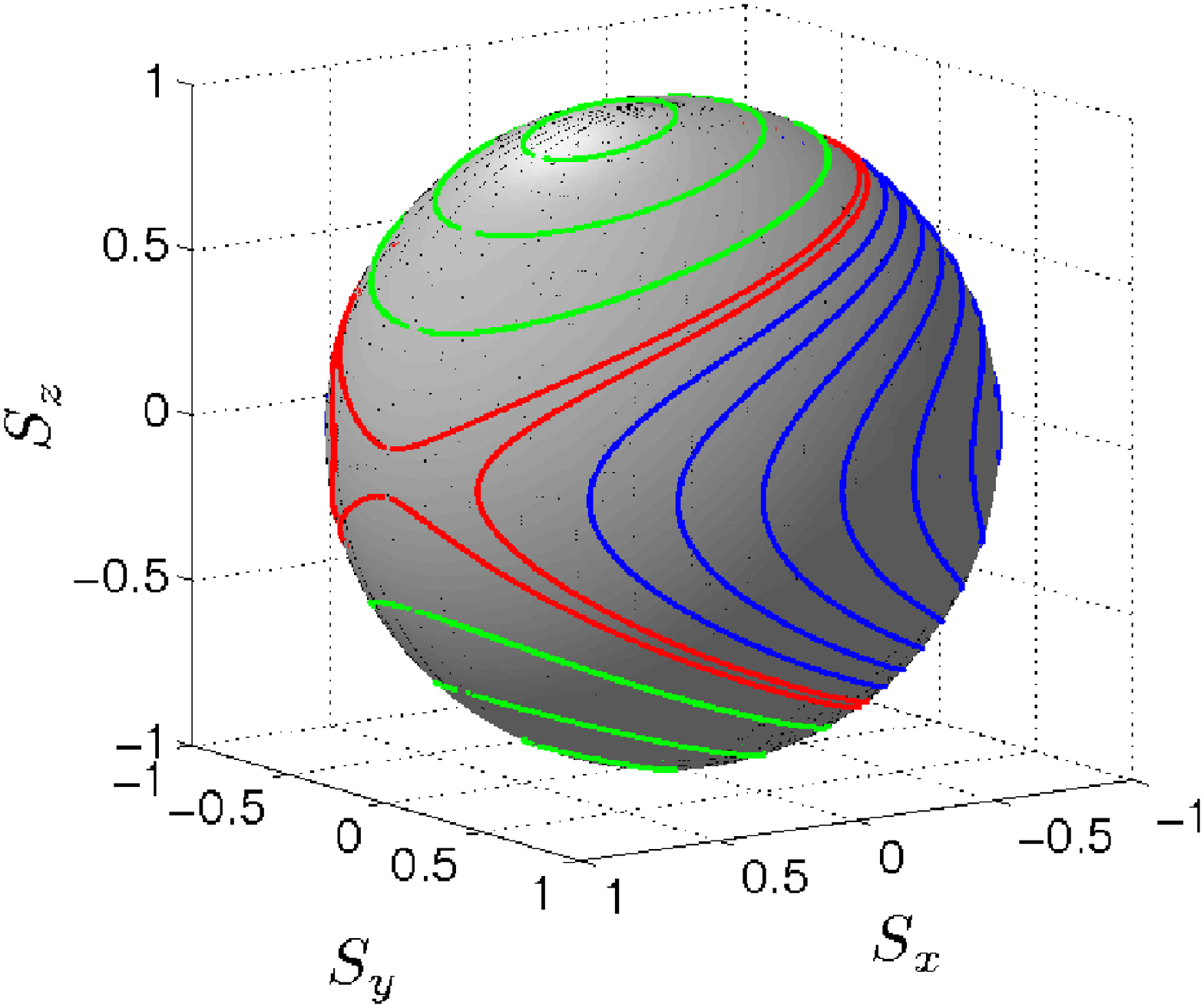} 
\includegraphics[width=7.5cm]{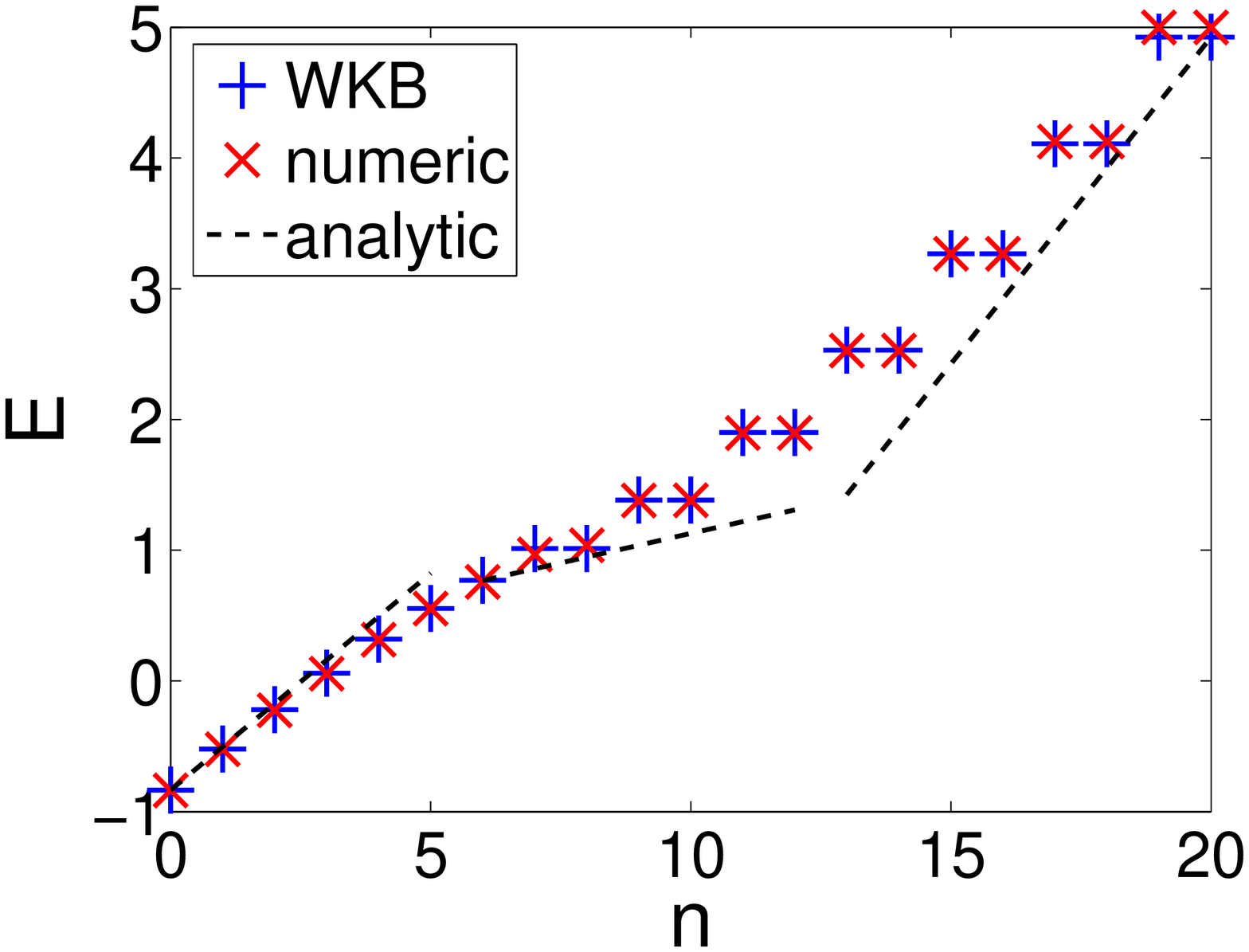} 

\caption{
(Color online)  
Contour lines for $u>2$. Sea levels are colored blue, Island levels are colored green, 
and the Separatrix is colored red (left panel).
Energy spectrum for $N={20}$ and $u=10$. 
WKB energies (red x) are compared with exact eigenvalues (blue +).   
Dashed lines indicate slopes  $\omega_J$ for low energies, 
$\omega_x$ for near-separatrix energies, 
and $\omega_{+}$ for high energies (right panel).
}

\label{f1}
\end{figure}

In \Fig{f1} we draw the constant energy contours ${\mathcal{H}(\theta,\varphi)=\const}$ of the top Hamiltonian (\ref{top}) 
for ${u=10}$ and ${\varepsilon=0}$. Generally, the qualitative features of the phase-space structure \cite{Vardi01,Mahmud05}, change drastically with the interaction strength. For ${u>1}$ a separatrix appears provided $|{\varepsilon}|<\varepsilon_c$, where  
\beq
\varepsilon_c \ \ = \ \ \ \left(u^{2/3}-1\right)^{3/2}~.
\eeq
This separatrix divides the spherical phase space into "sea" and two "islands" as shown in \Fig{f1}. In what follows, we focus on the case of zero bias. Accordingly we distinguish between 3~regimes
depending on the strength of the interaction \cite{Gati07}:
\beq
\mbox{Rabi regime:} && u<1~, \\
\mbox{Josephson regime:} && 1<u<N^2~, \\
\mbox{Fock regime:} && u>N^2~.
\eeq
In the Rabi regime the separatrix disappears and the entire phase-space consists of the nearly linear "sea". By contrast, in the Fock regimes the "sea" has area less than $1/N$, and therefore it cannot accommodate quantum states. Thus in the Fock regime phase space is composed 
entirely of two nonlinear components: the "islands" occupy the entire upper and lower hemispheres.  Our main interest below is in the intermediate Josephson regime 
where linear and non-linear regions coexist and the dynamics is least trivial. This is the regime of interest to most current BEC interferometers and luckily, precisely where semiclassical 
methods are expected to be most effective.

\section{WKB quantization and characteristic frequencies/level-spacings} 

Due to the simplicity of the two mode BHH (\ref{top}), it is possible to carry out its semiclassical quantization analytically \cite{Boukobza09a,SmithMannschott09,Boukobza09b,Franzosi00,Graefe07,Nissen10} and acquire great insight on the ensuing dynamics of the corresponding Wigner distribution \cite{wignerfunc,Agarwal81}. We begin by accurately determining the quantum energy levels in the Josephson regime, using the WKB prescription. Having a phase-space area of $4\pi$ spherical angle supporting $N+1$ quantum states, the Planck cell area is,  
\beq
h \ = \ \mbox{ {Planck cell area in steradians}} \ = \ \frac{4\pi}{N{+}1}~, 
\eeq
and the WKB quantization condition thus reads,  
\beq
A(E_{\nu}) \ \ = \ \ \left(\frac{1}{2} + \nu\right)h
\ \ \ \ \ \ \ \  {\nu=0,1,2,3,...}
\eeq
where $A(E)$ is the phase space area which is enclosed 
by the energy contour. Note that it does not matter 
which area, of which "side" of the contour, is selected.  
Away from the separatrix the levels spacing equals 
approximately to the classical oscillation frequency:     
\beq
\omega(E) \ \ \equiv \ \ \frac{dE}{d\nu} \ \ = \ \ \left[\frac{1}{h}A'(E)\right]^{-1}
\eeq
In particular in the absence of interaction this is the  Rabi frequency
\beq
 {\omega_K} \ \ \approx \ \  K 
\eeq
For strong interaction, 
in the bottom of the sea, 
it is the Josephson frequency
\beq
{\omega_J} \ \ \approx \ \  \sqrt{NUK} \ \ = \ \  \sqrt{u} \ \omega_K
\label{omegaJ}
\eeq
while in the top of the islands it is 
\beq
 {\omega_{+}} \ \ \approx \ \ \ \ \  NU  \ \ \ \ = \ \ \ u \ \omega_K
\label{omegaplus}
\eeq
Finally, in the vicinity of the separatrix 
a more careful analysis is required, leading 
to the $h$~dependent result  
\beq
{\omega_{\rm x}} \ \ \approx \ \ \left[\log\left(\frac{N^2}{u}\right)\right]^{-1}  {\omega_J}
\label{omegax}
\eeq
Comparing Eq.~(\ref{omegax}) to Eq.~(\ref{omegaJ}) and Eq.~(\ref{omegaplus}) we realize that only in the vicinity of the separatrix  does the number of particles ($N$) become an essential parameter in the 
spectral analysis of the dynamics at fixed $u$.

\section{The preparations} 

Current experiments in matter-wave interferometry enable the preparation of nearly coherent states of the $SU(2)$ algebra, by fast splitting of the condensate \cite{Schumm05}. Alternatively, number-squeezed states, approaching relative-number Fock states for large separation, can be prepared by slow, adiabatic splitting \cite{Est08}. So far, the main focus of study in the Josephson regime, has been on coherent population dynamics, contrasting coherent preparations located near the bottom of the linear sea which exhibit Josephson oscillations around the ground state \cite{Giovanazzi00,Cataliotti01,Albiez05,Levy07} , with coherent preparations located near the top of the nonlinear islands, which result in self-trapped phase-oscillations around the 'poles' \cite{Smerzi97,Albiez05}. Our focus here, is on the interesting effects incurred in the {\it fringe-visibility dynamics} of coherent preparations located {\it on the separatrix} and contrasting them with the more common ground-state, north-pole, and number-squeezed preparations.

\subsection{Wigner distributions}

In order to gain semiclassical insight, it is convenient to represent each eigenstates  $| E_{\nu} \rangle$ 
by a proper spin Wigner function \cite{wignerfunc,Agarwal81}, 
which is a quasi-distribution that dwells on the spherical phase space. 
In this representation these eigenstates corresponds to strips along the contour lines of $\mathcal{H}$. 
In the same representation Coherent states  {$|\theta\varphi\rangle$} 
are like a minimal Gaussian wavepackets, 
while Fock states {$|n\rangle$} are like equi-latitude annulus. 
Note that the coherent state $\theta{=}0$ is also a Fock state with all the particles occupying one site, while $|n=0\rangle$ is the Twin Fock state with equal number of particles in both sites.

\begin{figure}[h!]

\centering
\includegraphics[width=5cm]{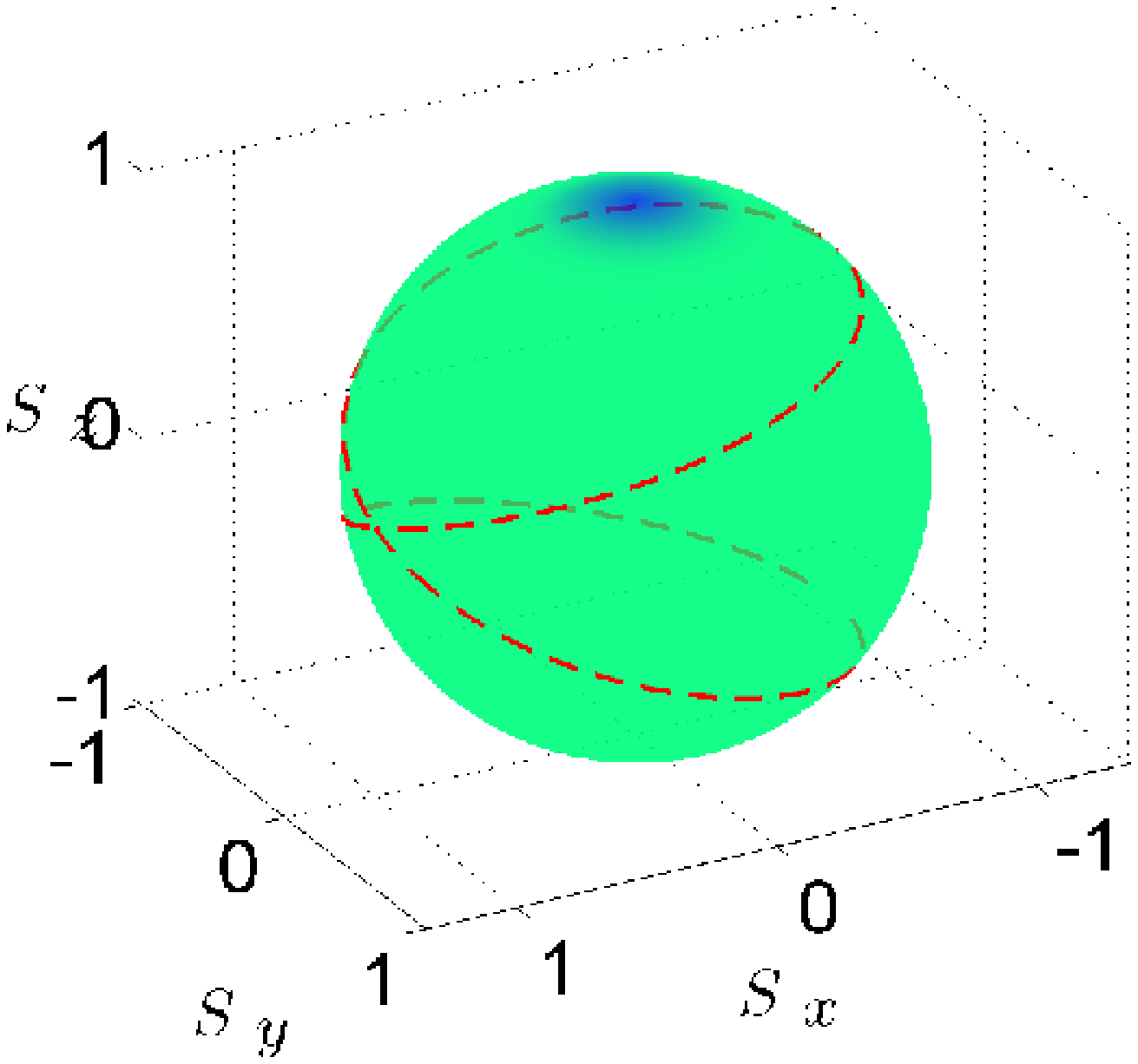} 
\includegraphics[width=5cm]{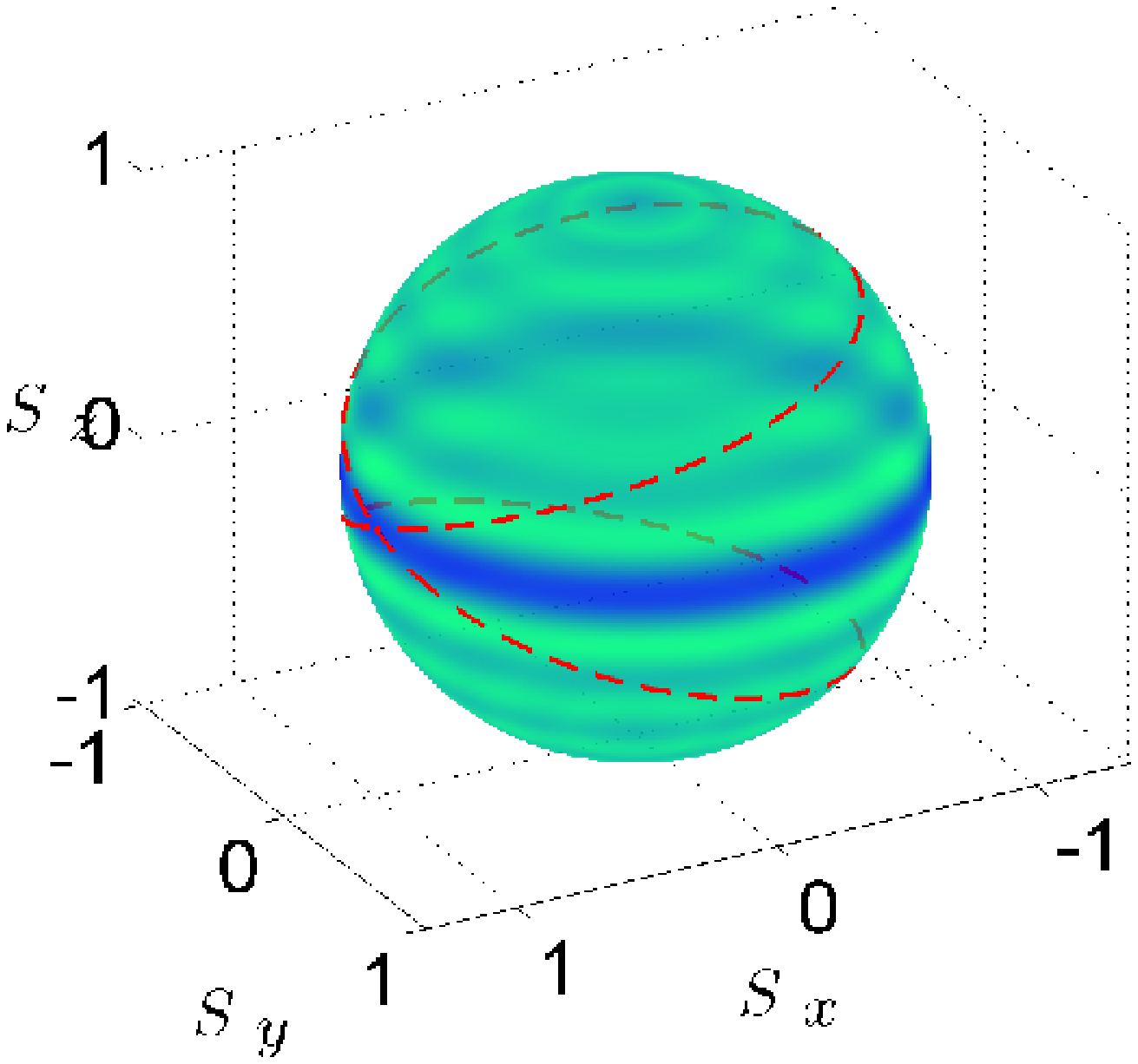}
\includegraphics[height=5cm]{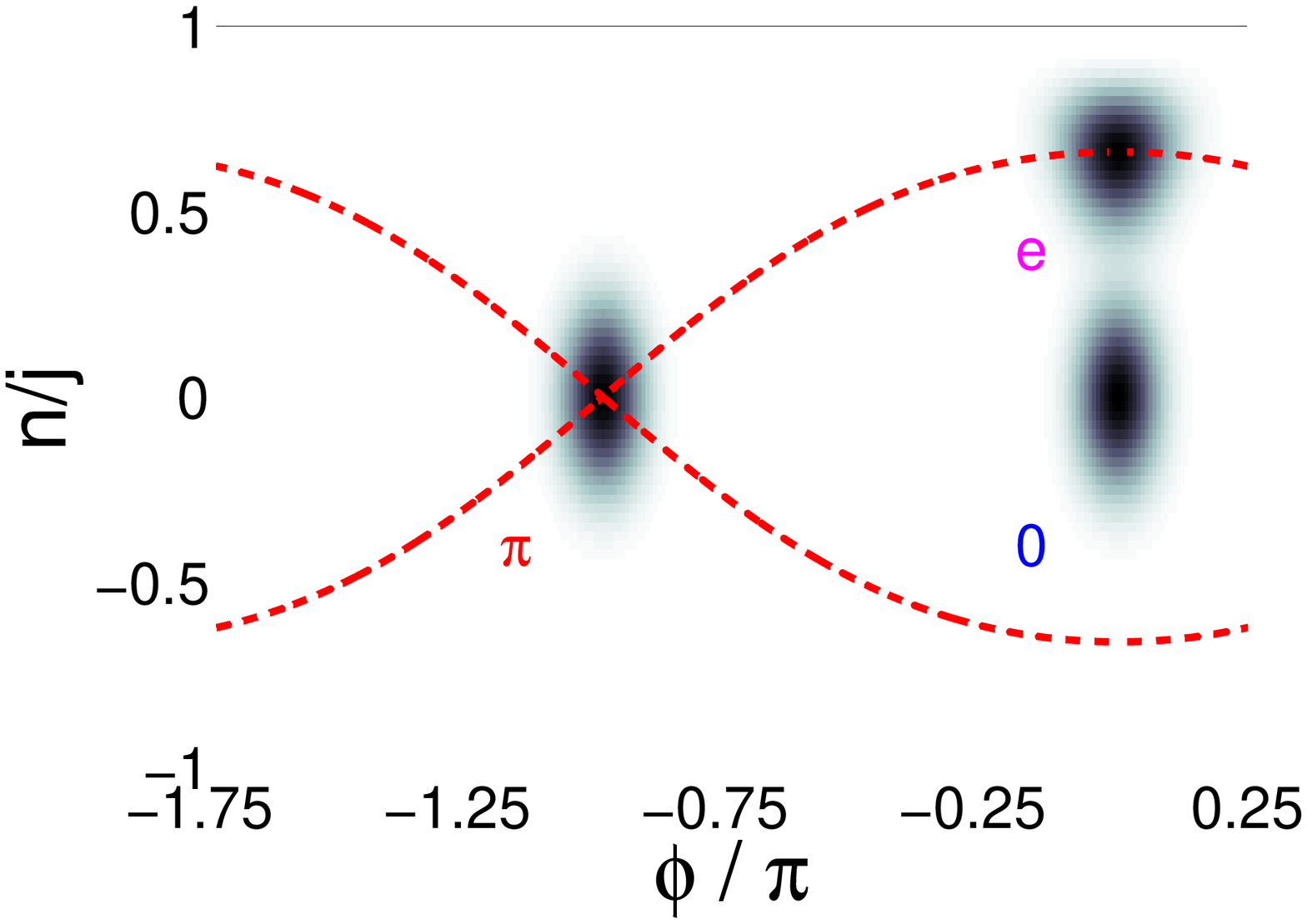} 

\caption{
(Color online) 
An illustration of the NorthPole ($\theta{=}0$) coherent state preparation (left), 
the TwinFock (${n=0}$) preparation (middle), 
and of Pi ("$\pi$"),  Zero ("$0$") and Edge ("e") preparations (right) 
using Wigner plots on a sphere. The left and middle panels are a 3D plots, 
while the right panel is a Mercator projection of the sphere 
using ${(\varphiJ,\nJ)}$ coordinates.
}
\label{f2}
\end{figure}
 
In \Fig{f2} we plot the Wigner functions corresponding to the five preparations under study. 
These include the NorthPole self-trapped state, the TwinFock state, two equal-population coherent states that we call Zero ($\varphi{=}0$) and Pi ($\varphi{=}\pi$),  and a third coherent state preparation that we call Edge. The two latter states (Pi and Edge) are both on-separatrix preparations. Note that in the Zero state all the particles occupy the symmetric orbital, 
while in the highly excited Pi state all the particles occupy the antisymmetric orbital.

Some of these preparations were experimentally realized and studied~\cite{Albiez05}.
While in the experimental work the emphasis was on contrasting Josephson-Oscillation and Self-Trapping,  with regard to near-Zero and near-NorthPole preparations, here our main interest is in contrasting the Zero preparation with the on-separatrix preparations Pi and Edge.

\subsection{Local Density of States}

\begin{figure}[h!]

\centering

\hspace{0.00\hsize} {\em TwinFock preparation} 
\hspace{0.08\hsize} {\em Zero preparation} 
\hspace{0.08\hsize} {\em Pi preparation} 
\hspace{0.12\hsize} {\em Edge preparation} 
\\
\includegraphics[width=0.24\hsize]{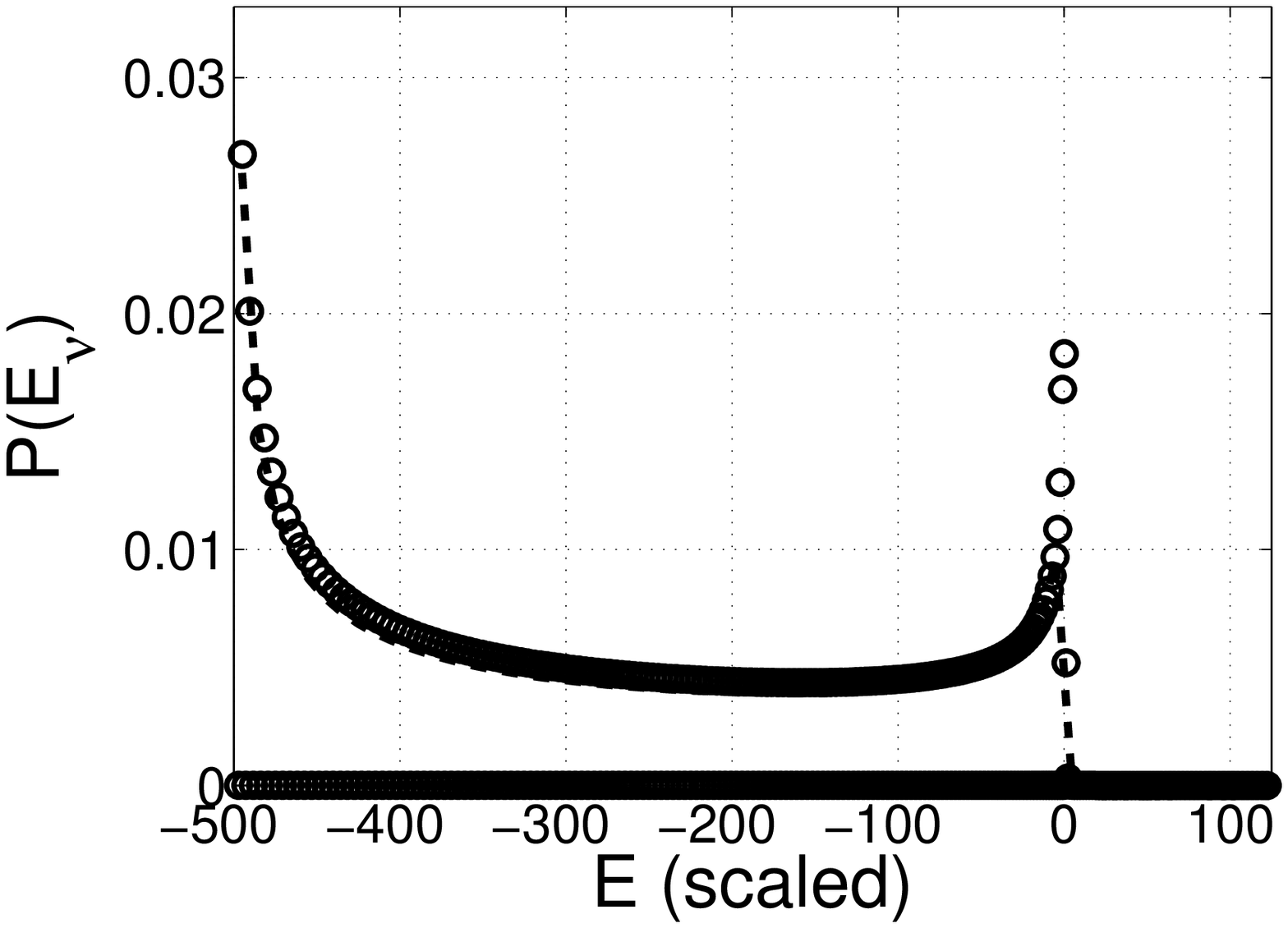} 
\includegraphics[width=0.24\hsize]{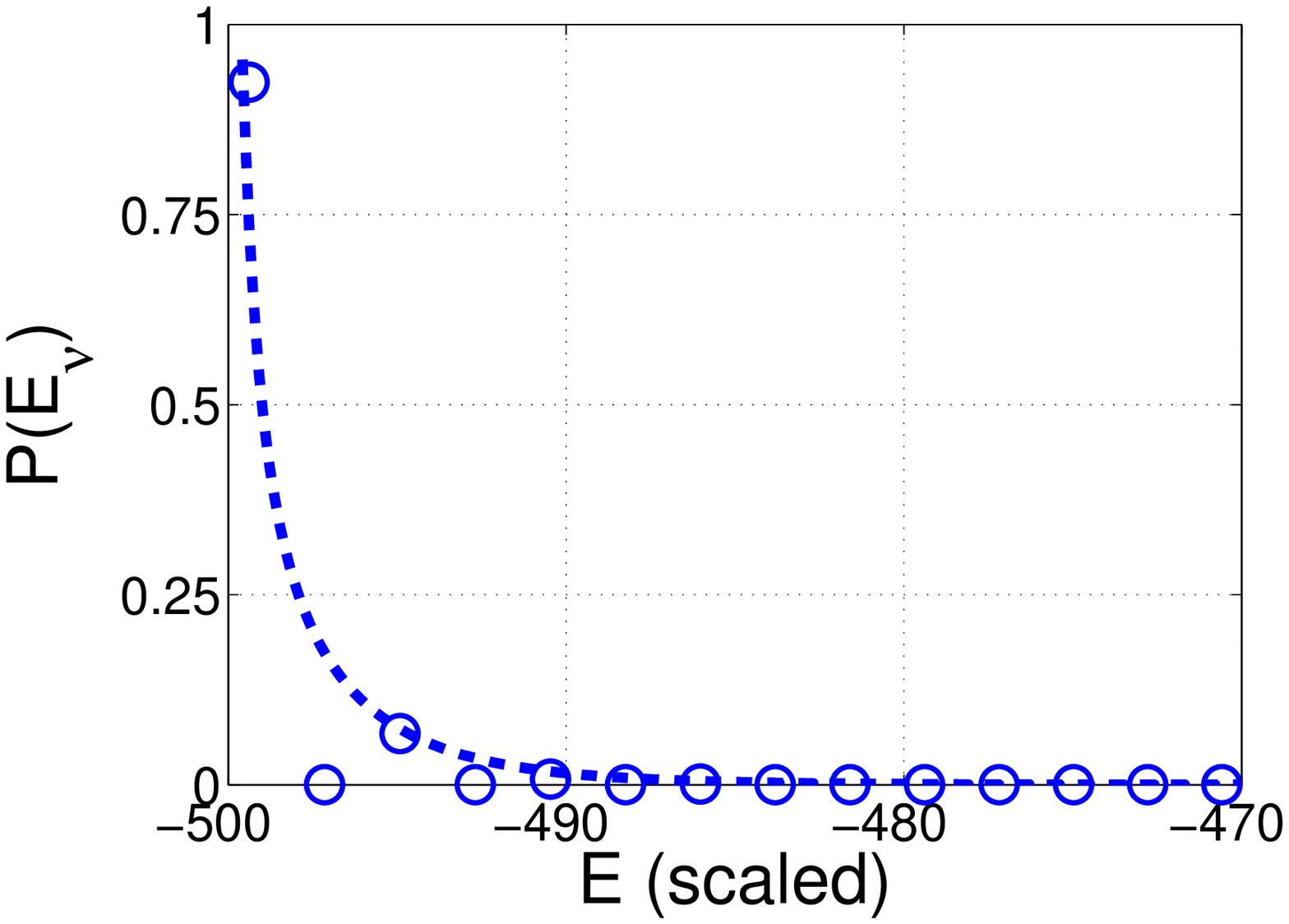} 
\includegraphics[width=0.24\hsize]{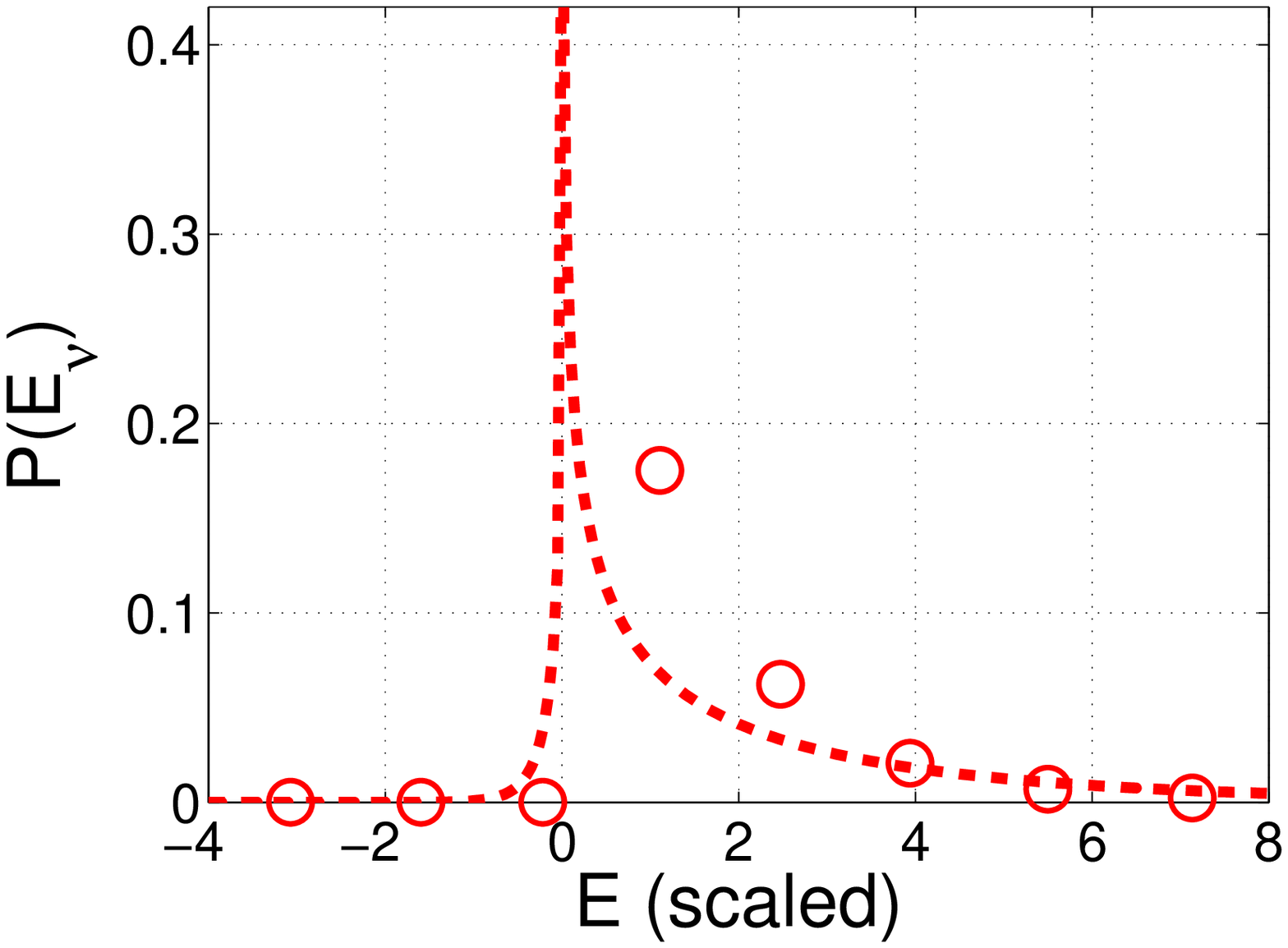} 
\includegraphics[width=0.24\hsize]{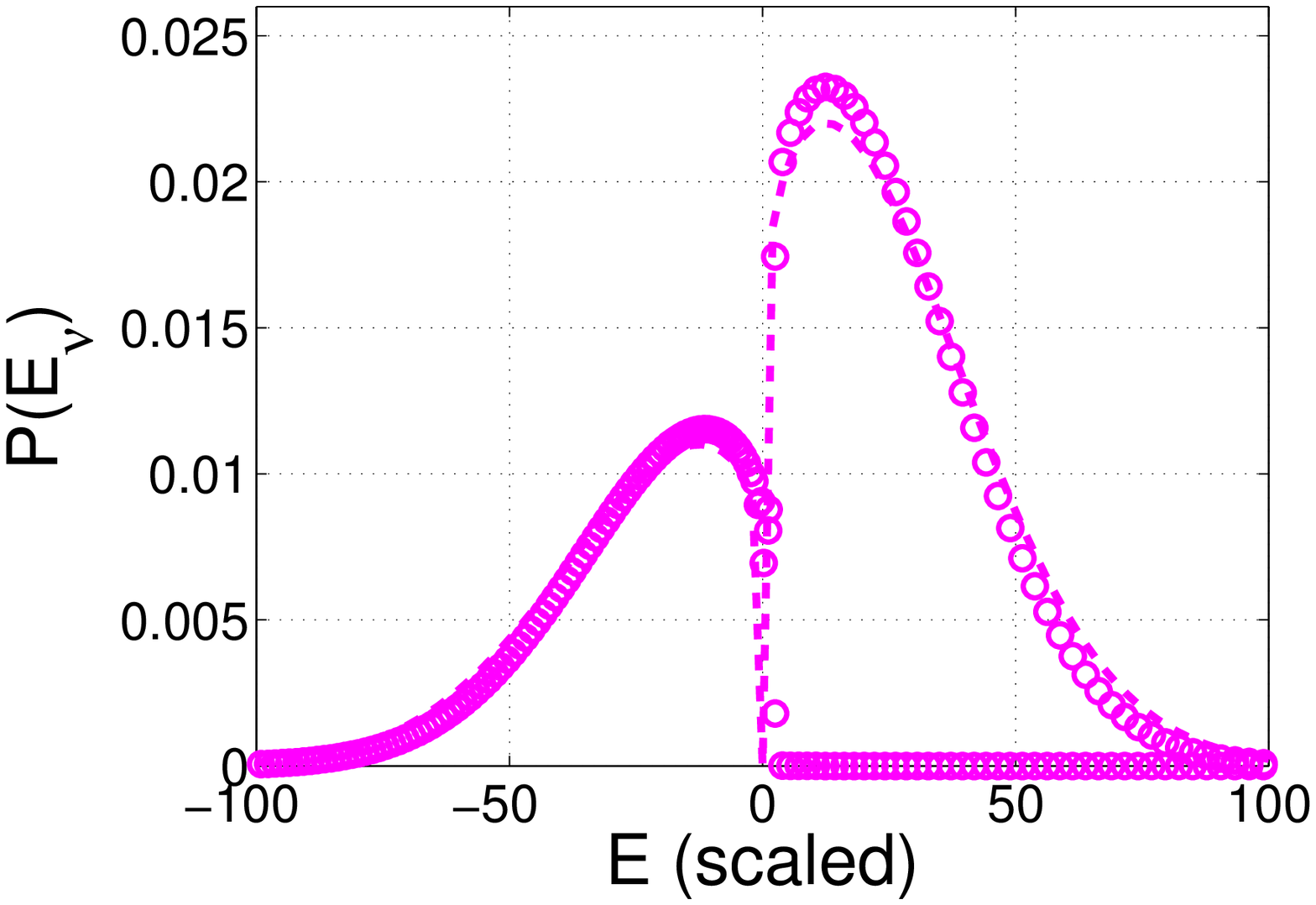} 

\caption{(Color online)
The LDOS of $N=500$  bosons with $u=4$, for TwinFock, Zero, Pi, and Edge preparations (left to right).
The horizontal axes are  $E-E_{\tbox{x}}$ and $\omega/\omega_J$.
The lines in the LDOS figures are based on a semiclassical analysis, 
while the circles are from the exact quantum calculation. 
Note the outstanding difference between the spectral support 
of Zero and Pi preparations compared with continuous-like support 
in the case of Edge and Fock preparations. 
}
\label{f5}
\end{figure}

In order to analyze the dynamics ensuing from the said preparations, we expand the initial
state as a superposition of the eigenstates $| E_{\nu} \rangle$.
The probability of the $\nu$th eigenstate in the superposition is denoted $\mbox{P}(E_{\nu})$, 
and is known as the local density of states (LDOS) with respect to the pertinent preparation. 
The LDOS of the various preparations is 
illustrated in \Fig{f5} and the line shape can be determined 
analytically via a semiclassical calculation \cite{Boukobza09a,Chuchem10}. Schematically 
the results can be summarized as follows:
\beq
\mbox{P}(E)\Big|_{\mbox{TwinFock}} \ \  &\sim& \ \ \left[1-\left(\frac{2E}{NK}\right)^2\right]^{-1/2} 
\\
\mbox{P}(E)\Big|_{\mbox{Zero}} \ \ &\sim& \ \ \mbox{\bf I} \left[\frac{E-E_{\tbox{-}}}{NU}\right]
\\
\mbox{P}(E)\Big|_{\mbox{Pi}} \ \ &\sim& \ \ \mbox{\bf K}\left[\frac{E-E_{\tbox{x}}}{NU}\right] 
\\
\mbox{P}(E)\Big|_{\mbox{Edge}} \ \ &\sim& \ \ \exp\left[-\frac{1}{N}\left(\frac{E-E_{\tbox{x}}}{\omega_J}\right)^2 \right]
\eeq
where ${\bf I}$ and ${\bf K}$ are Bessel functions. It is important 
to observe that the classical energy scales are $NK$ and $N^2U$. 
Accordingly only the line shape of the TwinFock LDOS has a purely  
classical interpretation. In contrast to that, the width of the 
coherent preparations is determined by the quantum uncertainty.

\subsection{Participation number} 

\begin{figure}[h!]
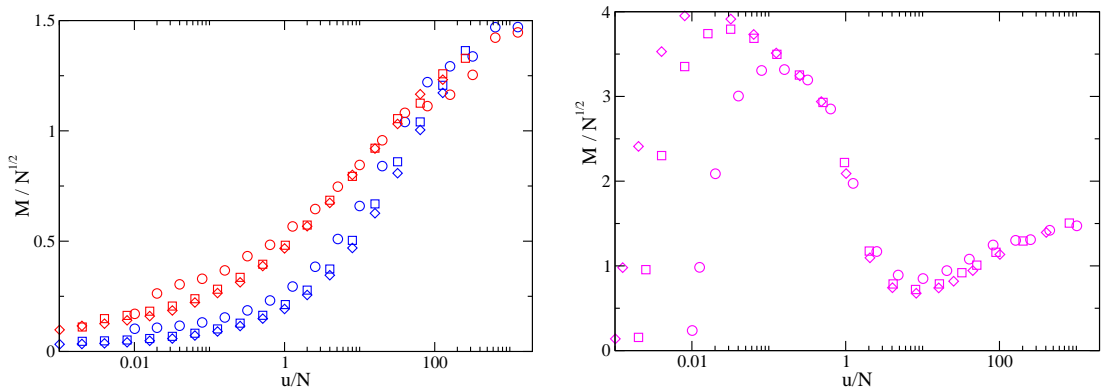


\centering
\includegraphics[clip,width=7cm]{M_phi_pi} 
\ \ \ \ 
\includegraphics[clip,width=7cm]{M_edge} 

\caption{(Color online) 
The participation number $M$ as determined from the LDOS  
for $N=100 \,(\circ),500\, (\Box),$ and $1000\,(\diamond)$ particles.
The left panel contains the Zero (lower set in blue) 
and Pi (upper set in red) preparations, 
while the Edge preparation is presented in the right panel.
Note the different vertical scale. 
In the crudest approximation we expect in the Edge case ${M \sim N^{1/2}}$, 
while in the Pi case  ${M \ll N^{1/2}}$ as long as ${(u/N)\ll 1}$  
}
\label{f6}
\end{figure}

The qualitative features of the fringe-visibility dynamics, given some initial preparation,  are determined by its {\it participation number}  defined as:
\beq
M \ \ \equiv \ \ \left[\sum_{\nu} \mbox{P}(E_{\nu})^2\right]^{-1} 
\ \ = \ \ \mbox{number of participating levels in the LDOS}
\eeq
In the case of a TwinFock preparation, 
the Wigner function is spread all over the equator of the spherical phase-space and thus overlaps with all the states in the sea up to the separatrix level (the equator intersects with all sea trajectories in Fig.~\ref{f2} but with no island trajectory).  Therefore we expect $M$ to be of order $N$, 
with classical ($N$~independent) prefactor 
that reflects the relative size of the sea:
\beq
M \ \ = \ \ \mbox{ClassicalPrefactor} \times N,  
\ \ \ \ \ \ \mbox{[TwinFock preparation]}
\eeq   
In the case of a coherent preparation, 
the Wigner function is a minimal wavepacket
that has width $\sigma_n=(N/2)^{1/2}$.
If the Fock states $| n \rangle$ were the eigenstates  
of the Hamiltonian, as they are in the Fock regime, we would get 
\be{101}
M \ \ = \ \ (2\pi N)^{1/2},  
\ \ \ \ \ \ \mbox{[Coherent preparation, $n$ basis]}~,
\eeq   
for all three coherent preparations. However in the Josephson regime the eigenstates 
are $| E_{\nu} \rangle$, and therefore the differences between the LDOS of the three coherent states Zero, Pi, and Edge,  are set by the ratio between $\sigma_n$ and the width of the separatrix (${\Delta n = NK/U}$).  The ratio $\sigma_n/\Delta n$ equals the 
dimensionless semiclassical parameter $(u/N)^{1/2}$. 
If this ratio is larger than unity the distinction 
between the Zero, the Pi and the Edge preparations 
is blurred, and we expect to have the same participation number.
Indeed plotting the participation number of these three initial states (\Fig{f6}), we see that at the strong interaction limit ${M\approx (3/2)N^{1/2}}$.
This is roughly half compared with \Eq{e101}, 
and reflects the odd-even selection rule
that removes half of the overlaps (the semiclassical states do not have a well-defined parity with respect to site substitution, whereas the actual quantum eigenstates are constructed from their odd and even superpositions. The participating constituents are those that have the same parity as that of the preparation).

For weaker interaction, when the semiclassical parameter $(u/N)^{1/2}$ is smaller than unity 
the different nature of the Zero, the Pi and the Edge preparations
expresses itself. Now we have to account for both the width 
of LDOS line shape and the mean level spacing. 
This leads to the following results:
\beq
M \ \ \approx \ \  \sqrt{u}, 
\ \ \ \ \ \ \ \ && [\mbox{Zero preparation}]
\\
M \ \ \approx \ \  \left[\log\left(\frac{N}{u}\right)\right] \sqrt{u}, 
\ \ \ \ \ \ \ \ && [\mbox{Pi preparation}]
\\
M \ \ \approx \ \  \left[\log\left(\frac{N}{u}\right)\right] \sqrt{N},
\ \ \ \ \ \ \  && [\mbox{Edge preparation}]
\eeq
The striking point here is that the Pi preparation resembles the Zero preparation, 
rather than its sister separatrix  Edge state. This seems at first site in contradiction
with semiclassical intuition: one would naively expect 
that wavepackets that have the same energy and reside in 
the same phase space region (separatrix) would behave 
similarly. This is not the case, as we see here, and later 
in the dynamical analysis. The Pi state is actually closer to the Zero preparation, 
as both have a small participation value. The resemblance  of the Pi and Edge preparations is only detectable in the formal limit $N\rightarrow\infty$.   In other words, because of the $N$ dependence of $M$ it is "easier" to approach  the "classical limit" in the case of an Edge preparation.

\section{Coherence dynamics} 

\subsection{Classical, semiclassical, and quantum dynamics of the Bloch vector}

\begin{figure}[h!]

\includegraphics[width=0.4\hsize]{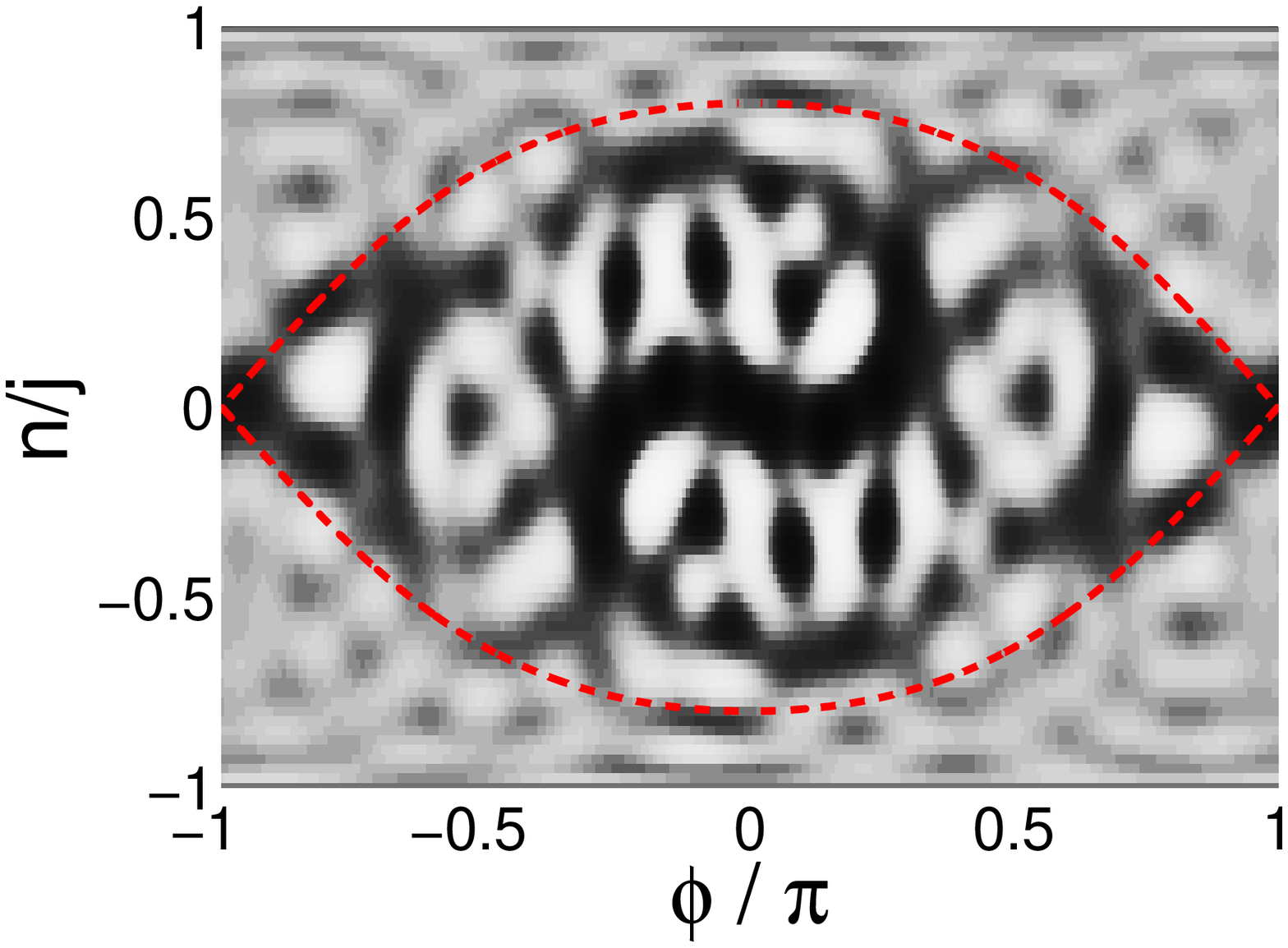}  
\includegraphics[width=0.4\hsize]{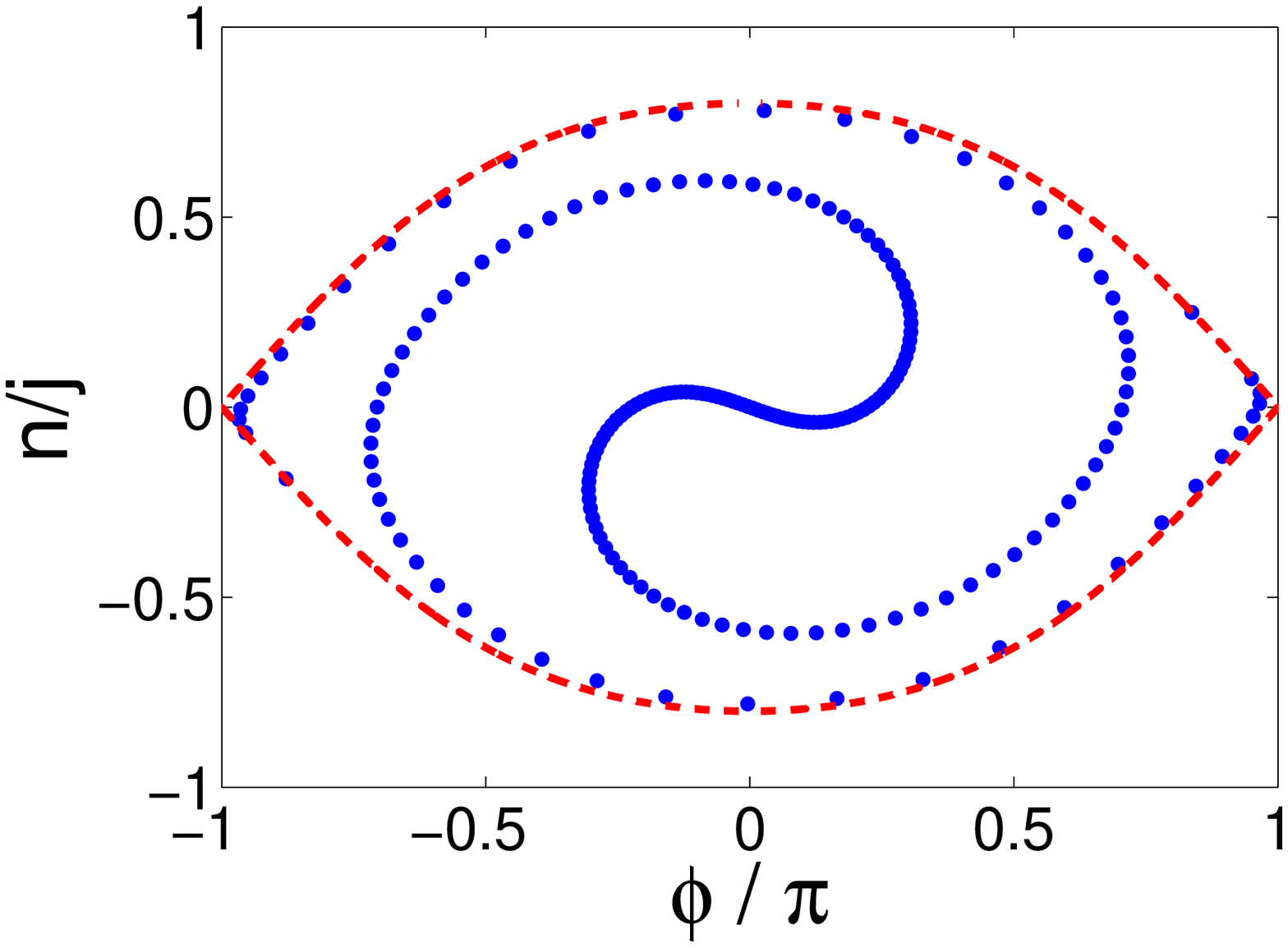} 

\caption{(Color online)
The evolving quantum state of $N=40$ bosons with $u=5$ for TwinFock ($n=0$) preparation.
The units are such that ${K=1}$ and the time is $t=4$. 
On the left - the Wigner function of the evolved quantum state.
On the right - the corresponding classical evolution.
}
\label{f3}
\end{figure}

In this section, we describe the dynamics of single-particle coherence for the preparations of Section IV. The lowest-order approximation is the {\it mean field} dynamics.  It is generated by replacing the operators in the Hamiltonians of Eq.~(\ref{BHH}), Eq.~(\ref{spin}), or Eq.~(\ref{top}) by $c$-numbers, thus obtaining a set of classical equations of motion for them. For example, in the spin representation, 
\begin{eqnarray}
{\dot J}_x&=&(\mathcal{E}-2U J_z) J_y~,\nonumber\\
\label{heisen}
{\dot J}_y&=&K Jz - (\mathcal{E}-2U J_z) J_x~, \nonumber\\
{\dot J}_z&=&-KJ_y~.
\label{MFT}
\end{eqnarray}
These Gross-Pitaevskii equations (GPE) describe the {\em classical} evolution  of a {\em point} in phase space, which is a single trajectory. The classical evolution assumes that the state of the system is coherent at all times, so that the single-particle coherence is fixed to unity and the Wigner distribution always resembles a minimal Gaussian (the center of which is the traced point). By contrast, the {\em semi-classical} theory describes the classical evolution and subsequent deformation of a {\em distribution} in phase space, according to the GPE equations (\ref{MFT}).  Finally the {\em Quantum theory} is obtained by direct solution of the Schr\"odinger or Heisenberg equations with the Hamiltonians (\ref{BHH}),  (\ref{spin}), or (\ref{top}). This full quantum solution adds recurrences and fluctuations which are absent from the classical and semiclassical pictures and result from the discreteness of the energy spectrum. The dimer 
system is integrable, and therefore the WKB method provides 
a very good basis for the analysis. \Fig{f3} illustrates the agreement
between the quantum evolution of the Wigner function, starting from the TwinFock state \cite{Boukobza09b,Chuchem10} and the semiclassical 
evolution of a corresponding distribution. 
It should be emphasized that in the Wigner-Wyle formalism 
any operator $\hat{A}$ is presented by the phase-space 
function  $A_{\tbox{W}}(\Omega)$, and the calculation 
of an expectation value can be done in a classical-like formulation: 
\beq
\bra \hat{A} \ket \ \ = \ \ 
\trc[\hat{\rho} \ \hat{A} ] \ \ = \ \ 
\int\frac{d\Omega}{h}{\rho_{\tbox{W}}(\Omega) A_{\tbox{W}}(\Omega) }
\eeq

\begin{figure}[h!]

\centering
\includegraphics[width=0.31\hsize]{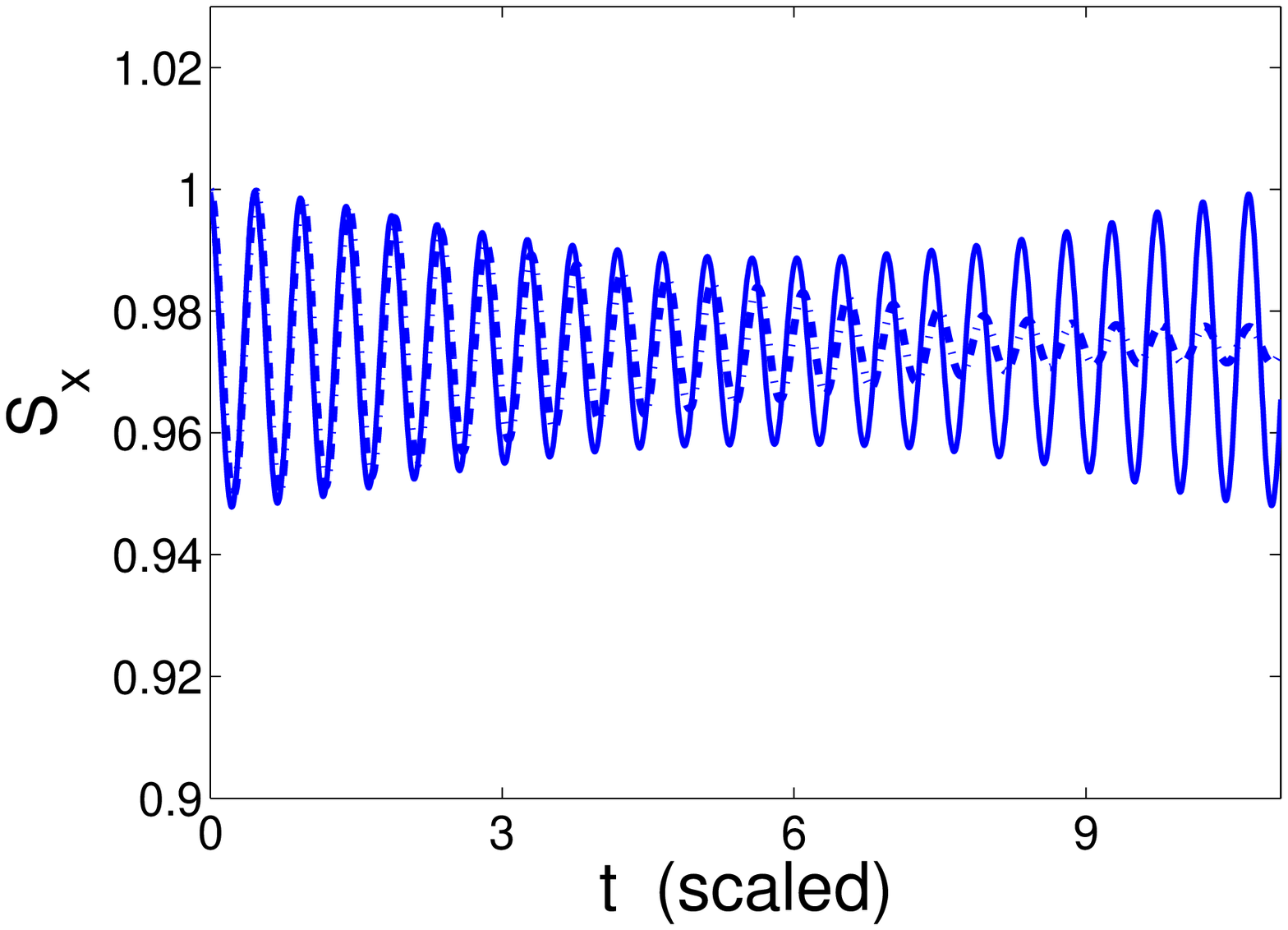}
\includegraphics[width=0.31\hsize]{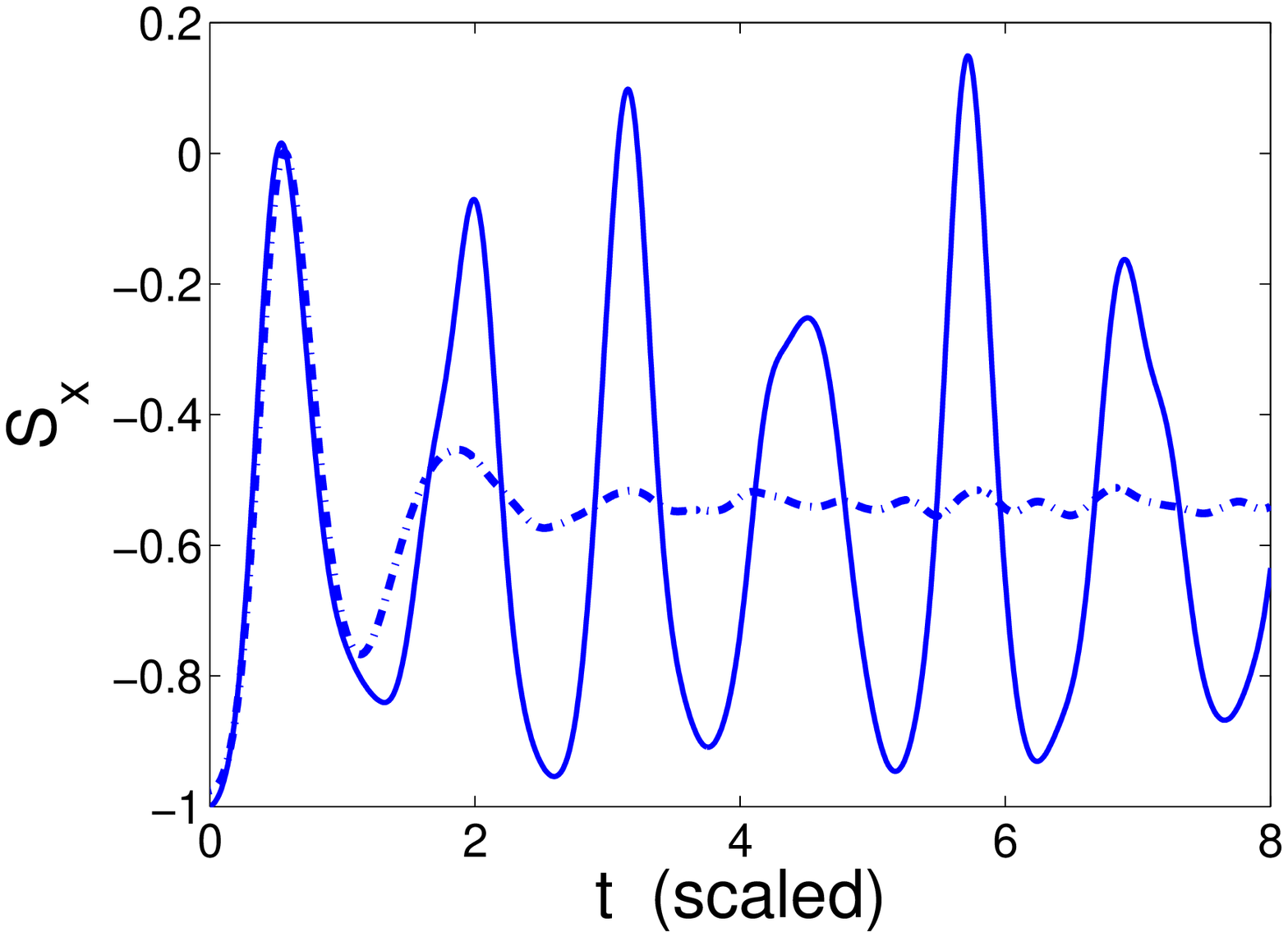}
\includegraphics[width=0.31\hsize]{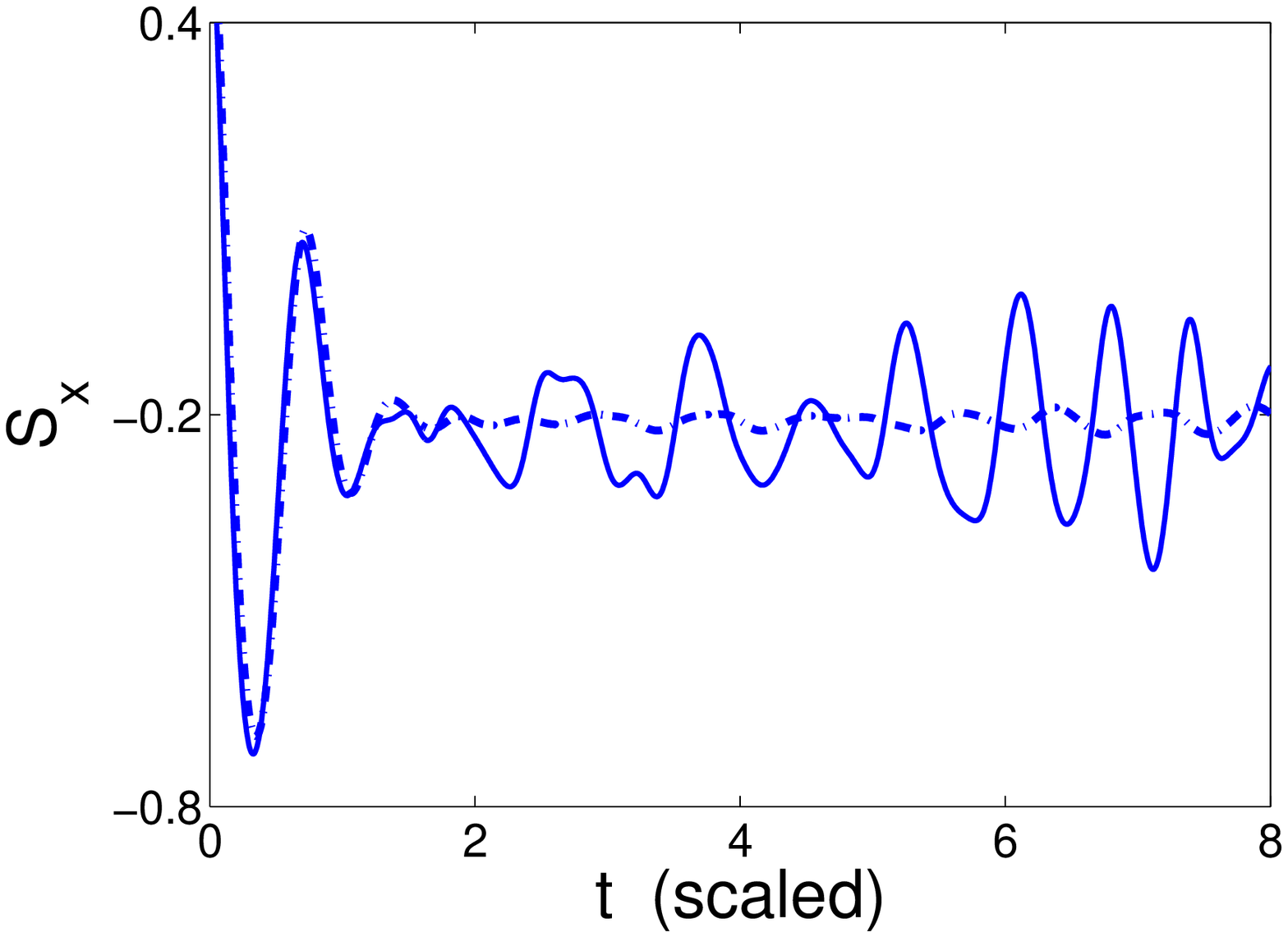}

\caption{(Color online) 
The variation of $S_x(t)$ with time for $N{=}40$ particles with $u=5$,   
for Zero (left), Pi (middle), and Edge (right) preparations. 
Note the different vertical scale. 
The dashed-dotted lines are based on semiclassical simulation.
}
\label{f4}
\end{figure}

The single-particle density matrix of the two-mode system can be presented using the Bloch vector, 
\beq
\vec{\bm{S}} = \langle\vec{\bm{J}}\rangle  / (N/2)  &=&   {(S_x, S_y, S_z)} ~,
\eeq
Whereas previous work \cite{Albiez05} has focused on the mean-field dynamics of the occupation difference $(N/2)  \ \langle S_z \rangle$, here we study the {\it single-particle coherence} manifested in the one body purity, 
\beq
\mbox{OneBodyPurity} &=&  (1/2) \left[ 1 + \langle S_x \rangle^2 + \langle S_y \rangle^2 + \langle S_z \rangle^2 \right]~,
\eeq
and the transverse component of the Bloch vector, the fringe visibility,
\beq
g_{12}^{(1)} &=&  \left[ \langle S_x \rangle^2 + \langle S_y \rangle^2 \right]^{1/2}~.
\eeq
This quantity reflects the fringe-contrast over multiple runs of an experiment in which the particles from the two confined modes are released and allowed to interfere.
In \Fig{f4} we plot examples for the evolution of the Bloch vector, 
and observe significant differences in the dynamical behavior of the Zero, Pi, and Edge preparations. 
Our objective is to understand how the dynamics depends 
on the dimensional parameters: the "classical" parameter~$u$ 
and the "quantum" parameter~$N$. 

In accordance with the opening paragraph of this section, 
we see in \Fig{f4} that in the semiclassical simulation the fluctuations always 
die after a transient. This should be contrasted with both 
the {\em classical} (single trajectory) behavior, 
and the {\em quantum} behavior. 
In the latter quantum case the wavepacket 
is a superposition of ${M>1}$ eigenstates, 
and consequently there are persistent fluctuations 
that depend on the "quantum" parameter~$N$.  In what follows, we will quantitatively analyze the characteristic features of the quantum dynamics resulting from the three coherent preparations, including the frequency of oscillation, its mean long-time value, and its RMS amplitude, and compare the analytic predictions to the numerical results at various values of the characteristic parameter $u/N$ .

\subsection{Characteristic frequencies}

\begin{figure}[h!]

\centering
\includegraphics[clip,width=0.4\textwidth]{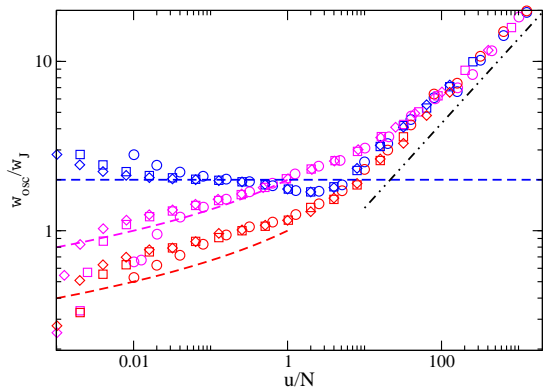}  

\caption{(Color online)
The mean frequency of the $S_x(t)$ oscillations versus $u/N$ for
for $N=100 \,(\circ),500\, (\Box),$ and $1000\,(\diamond)$ particles.
The preparations are (upper to to lower sets of data points):  
Zero (blue), Edge (magenta), and Pi (red).
The weaker-interaction theoretical predictions (\ref{wzero})-(\ref{wedge}),
which are doubled due to mirror symmetry,  
are represented by blue, red, and magenta dashed lines, 
while the strong interaction prediction (\ref{wstrong}) is 
represented by a black dash-double-dotted line.
}
\label{f7}
\end{figure}

The typical frequency of the fluctuations is the simplest characteristic 
that differentiate the three panels of \Fig{f4}. 
The numerically-obtained frequency~$\omega_{\tbox{osc}}$ as a function 
of the interaction parameter~$u/N$ is displayed in~\Fig{f7}. 
In the classical picture $\omega_{\tbox{osc}}$ 
should be related to the Josephson frequency $\omega_J$ at the bottom of the sea, 
while quantum mechanically it reflects the level spacings  
of the participating levels. A straightforward analysis 
leads to the followings estimates:
\beq
\label{wzero}
\omega_{\tbox{osc}} \ \  \approx & \ \ 
2 \omega_J
&  \ \ \ \ \ \ \ \ \ \ \ \ \ \ \  {[\mbox{Zero}]}  
\\
\label{wpi}
\omega_{\tbox{osc}} \ \  \approx & \ \ 
 {1\times} \left[\log\left(\frac{N}{u}\right)\right]^{-1} \ 2\omega_J
&  \ \ \ \ \ \ \ \ \ \ \ \ \ \ \  {[\mbox{Pi}]}
\\
\label{wedge}
\omega_{\tbox{osc}} \ \  \approx & \ \ 
 {2\times} \left[\log\left(\frac{N}{u}\right)\right]^{-1} \ 2\omega_J
&  \ \ \ \ \ \ \ \ \ \ \ \ \ \ \  {[\mbox{Edge}]}  
\\
\label{wstrong}
\omega_{\tbox{osc}} \ \  \approx & \ \ 
\left(\frac{u}{N}\right)^{1/2} \ 2\omega_J
& \ \ \ \ \ \ \ \ \ \ \ \ \ \ \ [ u \gg N]
\eeq
The first three expressions apply for ${(u/N)<1}$, 
where the differences between the preparations is distinct.
The last expression refers to the regime ${(u/N)>1}$
where the differences are blurred and the three preparations become equivalent equatorial states 
(as clearly evident from \Fig{f7}).
Note that due to the mirror symmetry of the Zero preparation  
the expected frequency should approach $2\omega_J$, 
while for the Pi preparation it is bound from below by $2\omega_{\tbox{x}}$. 
Both frequencies are indicated in \Fig{f7} by dashed lines.

\subsection{Long time average} 
\begin{figure}[h!]
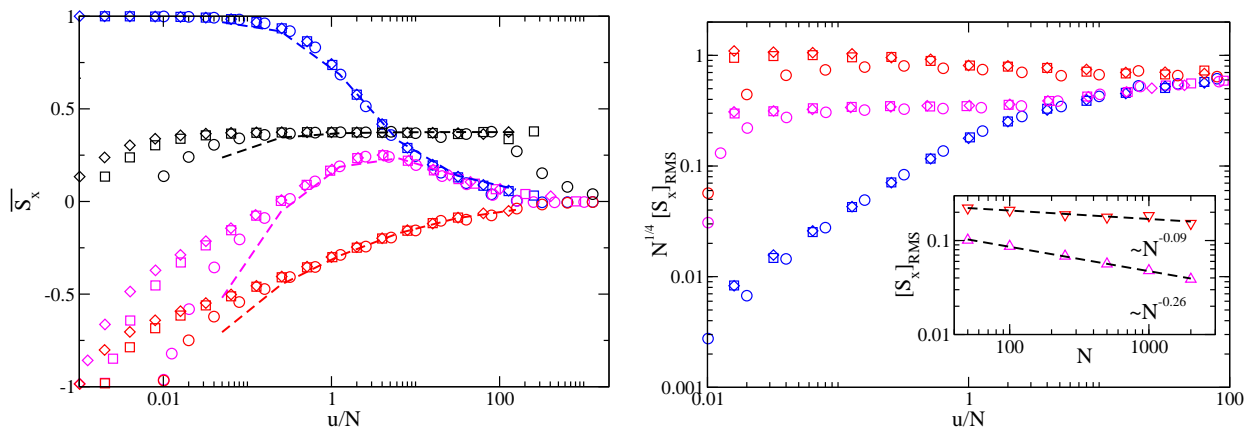


\centering
\includegraphics[clip,width=8cm]{s_x}
\ \ \ 
\includegraphics[clip,width=8cm]{RMS_s_x_scaled} 

\caption{(Color online)
{\em Left:} The long-time average of $S_x(t)$ versus $u/N$
for $N=100 \,(\circ),500\, (\Box),$ and $1000\,(\diamond)$ particles.
The preparations are (upper to lower sets of data points):  
Zero (blue), TwinFock (black), Edge (magenta), and Pi (red). 
The symbols are used for the quantum results and
the dashed lines are the semiclassical prediction for fourty particles.
Note that the scaling holds only in the Josephson regime ${1\ll u \ll N^2}$,
and therefore, for a given $u/N$ range, becomes better for large~$N$.
{\em Right:} 
The long time RMS of $S_x(t)$ for the three coherent preparations 
(lower to upper sets): Zero (blue), Edge (magenta), and Pi (red). 
In the inset, the RMS of $S_x(t)$ for Edge ($\triangle$) and Pi ($\triangledown$) preparations
is plotted versus~$N$ while~${u=4}$ is fixed.
}    

\label{f8}
\end{figure}

Next we examine the long time average, plotted in \Fig{f8} as a function of the characteristic parameter.  Unlike the fluctuations around it, this average value does not reflect the quantization of energy and therefore a purely semiclassical analysis is adequate. 
The naive expectation might be that coherence would be diminished due to   
phase spreading (aka "phase diffusion"). This is indeed the case in the  Fock regime, 
leading to $\langle S_x \rangle_{\infty} \approx 0$. However, the situation is rather more complicated in the  Josephson regime, where $\langle S_x \rangle_{\infty}$ is determined by $u/N$.
The semi-classical phase space picture allows to calculate  
the phase distribution $\mbox{P}(\varphi)$ that pertains  
to the long time ergodic-like distribution (see for example \Fig{f3}). 
This distribution is determined by the LDOS. Then we use the integral 
\beq
\overline{S_x} \approx \int \cos(\varphi) \, \mbox{P}(\varphi) d\varphi~,
\eeq
to evaluate the residual coherence. This procedure results in the following predictions:   
\beq
\label{TFcoherence}
\overline{ S_x } \approx
& 1/3 \ \ \ 
& \ \ \ \ \  {\mbox{[TwinFock]}}
\\
\overline{ S_x } \approx
& \ \ \ \exp[-(u/N)]  
& \ \ \ \ \  {\mbox{[Zero]}}
\\ 
\overline{ S_x } \approx
& -1-4/\log\left[\frac{1}{32}(u/N)\right]  
& \ \ \ \ \  {\mbox{[Pi]}}
\eeq
Thus, the coherence of the Zero preparation is robustly maintained as long as  {$u/N<1$}, 
corresponding to the $\varphi=0$ phase locking of the two condensates due to the weak coupling. 
By contrast, the Pi and Edge coherence is far more fragile throughout the Josephson regime \cite{Boukobza09a}.
Note that the phase locking of the Pi state would take place only if it were the Rabi-regime (no separatrix).    
As evident from the Twin-Fock self-induced coherence (\ref{TFcoherence}), 
one-particle coherence should not necessarily be lost due to interactions, 
but could actually be {\it built} \cite{Boukobza09b}. 
This course of events is somewhat similar to the coherent relaxation of a system 
to its ground state at low temperatures: the ground state has higher purity compared with the initial preparation.

\subsection{RMS of the fluctuations} 
Finally, we turn to discuss the RMS of the fluctuations, 
which constitute a fingerprint of energy quantization. 
General reasoning implies that the {\em classical} 
fluctuations are suppressed by factor~$M$:  
\be{73}
\mbox{RMS}\left[\bra A \ket_t \right] 
= \left[\frac{1}{M} \int \tilde{C}\tbox{cl}(\omega) d\omega\right]^{1/2} 
\eeq
In this formula $\tilde{C}\tbox{cl}(\omega)$ is the {\em classical} 
power spectrum of an ergodic trajectory. What we want to highlight 
is the {\em quantum} $N$~dependence. It is important to clarify that 
in the {\em semi-classical} limit $M\rightarrow\infty$, and therefore 
the fluctuations are suppressed. We emphasize again that the quantum 
behavior is intermediate between the coherence-preserving {\em classical} (single trajectory) 
dynamics and the strong coherence-attenuation of the  {\em semi-classical} (infinite $M$) dynamics.

The dependence of~$M$ on~$N$ is remarkably different for the 
various preparations. In the case of the TwinFock preparation ${M\sim N}$
and therefore the RMS is inversely proportional to $N^{1/2}$. 
This should be contrasted with the case of the Pi and the Edge preparations:
\be{74}
\mbox{RMS}\left[ S_x(t) \right] 
 \sim &  
N^{-1/2}   
& \ \ \ \ \  {[\mbox{ TwinFock}]}  
\\
\mbox{RMS}\left[ S_x(t) \right] 
 \sim &  
N^{-1/4}   
& \ \ \ \ \  {[\mbox{ Edge}]}  
\\
\mbox{RMS}\left[ S_x(t) \right] 
 \sim &  
(\log(N))^{-1/2}  
& \ \ \ \ \  {[\mbox{ Pi}]}
\eeq
We further note that in the Pi case the leading 
dependence of the participation number is
on the classical parameter ($M\sim u^{1/2}$), 
unlike the case of the Edge preparation where 
the leading dependence in on the quantum parameter ($M\sim N^{1/2}$).   
The results of the RMS analysis are presented in \Fig{f8}.
The implied $N^{1/4}$ scaling based on Eq.~(\ref{e73})) is confirmed.  
The dashed lines in the inset are power-law fits 
that nicely agree with the predictions of Eq.~(\ref{e74}).

\section{Summary} 

Using semiclassical machinery, we have analyzed the temporal fluctuation of the single-particle coherence, and of the fringe-visibility, in the Bose Josephson model. While recent experiments in the Josephson regime have essentially focused on mean-field population dynamics \cite{Albiez05}, with coherent preparations at the spectral 
extremes (self trapping versus Josephson oscillations), here we highlight intricate effects that can be found 
by studying the coherence-dynamics in the intermediate separatrix regime. We predict significant differences 
in the transverse relaxation of seemingly similar coherent initial states, differing by the initial relative-phase 
or by their location along the separatrix, as well as the interaction-induced phase-locking of two initially separated BECs due to the combined effect of interaction and coupling.  
The semiclassical WKB quantization facilitates the calculation of the LDOS for the pertinent preparations, 
and thus the estimation of the number of participating eigenstates. This allows for a detailed quantitative analysis 
of the time evolution of the fringe visibility function. The obtained analytic expressions are found to be in a very good agreement 
with the results of numerical calculations.

\ \\

{\bf Acknowledgments.-- } 
The current manuscript summarizes the approach taken in a series of recent 
publications \cite{Boukobza09a,Boukobza09b,SmithMannschott09,Chuchem10}. 
We acknowledge contributions to these works by Maya Chuchem, Erez Boukobza, 
Katrina Smith-Mannschott, Moritz Hiller, and Tsampikos Kottos.
This research was supported by the Israel Science Foundation (grant Nos. 29/11 and 346/11) and by Grant No. 2008141 from the United States-Israel Binational Science Foundation (BSF) .


\end{document}